\documentclass[preprint,preprintnumbers,amsmath,amssymb]{revtex4}
\usepackage[dvips]{graphicx}
\usepackage{psfrag}
\usepackage{rotating}
\usepackage{amsmath}
\usepackage{amsfonts}
\usepackage{amssymb}

\newcommand{\al}{\alpha}
\newcommand{\non}{\nonumber}

\begin{document}


\title{About non-existence of the molecular ion $\bf H_3^{++}$}

\author{H. Medel-Cobaxin}
\email{medel@nucleares.unam.mx}
\affiliation{Instituto de Ciencias Nucleares, UNAM, Apartado Postal
  70-543, 04510 M\'exico}
\author{A. Alijah}
\email{alijah@ci.uc.pt}
\affiliation{Departamento de Qu\'{\i}mica, Universidade de Coimbra, \\
3004-535 Coimbra, Portugal}
\author{A.V. Turbiner}
\email{turbiner@nucleares.unam.mx}
\affiliation{Instituto de Ciencias Nucleares, UNAM, Apartado Postal
  70-543, 04510 M\'exico ~}
\begin{abstract}
  It is shown that the molecular ion ${\rm H}^{++}_3$ does not exist
  in a form of the equilateral triangle. To this end, a compact variational
  method is presented which is based on a linear superposition of six specially
  tailored trial functions containing non-linear parameters. Careful optimization
  of a total of fifteen parameters gives consistently lower variational results for
  the electronic energy than can be obtained with standard methods of quantum
  chemistry even with huge basis sets as large as mcc-pV7Z.
\end{abstract}

\begin{center}
{\it \large Dedicated to Professor Rudolf Zahradnik on
the occasion of his 80th birthday}
\end{center}

\maketitle

\section{Introduction}

The trihydrogen dication, $\rm H_3^{++}$, which consists of three protons
and one electron, is among the simplest Coulomb systems. Its stability
has been studied intensely in the sixties and early seventies.
In a series of articles, Conroy~\cite{conroy1, conroy2, conroy3} investigated the potential energy surfaces of the electronic ground state and the lowest excited states at linear and isosceles triangular configurations. He employed a variational approach in which the electronic trial wavefunction is expanded around the center of the nuclear charges. Analyzing the contour plots Conroy concluded that $\rm H_3^{++}$ is not stable.  Schwartz and Schaad \cite{schwartz}, and Somorjai and Yue~\cite{somorjai}, who reported single-point calculations of the system $(pppe)$ at the supposed equilibrium equilateral triangular configuration of $\rm H_3^+ $, did not address the stability problem. To assess Conroy's results, Berkowitz and Stocker~\cite{berkowitz} searched for this ion through charge stripping experiments on $\rm H_3^+$. They could not find evidence of stable $\rm H_3^{++}$. Later, the issue was reconsidered also from the theoretical side,
by Shoucri and Darling~\cite{darling}, who examined equilateral
configurations with the variational linear combination of atomic orbitals (LCAO) method, and by Hern\'andes and Carb\'o~\cite{carbo},
who studied two particular configurations with a more compact variational approach and obtained total energy values below those published before. No bound state has been determined in these calculations. Johnson and Poshusta~\cite{johnson} reported another single-point calculation in the context of Gaussian basis set optimization for some one-electron systems.

About twenty years later Ackermann {\em et al.}~\cite{ackermann} revisited
the question about the existence of $\rm H_3^{++}$ using the finite element method which provided much higher accuracy than previously achieved.
The problem of the stability of $\rm H_3^{++}$ was treated keeping the nuclear charge as a continuous parameter. Critical values of the charges for the existence of stable or metastable  equilateral triangular configurations were
obtained as $Z_c^+= 0.82$ and $Z_c^+= 0.95$, respectively. The authors
excluded the possibility of stable $\rm H_3^{++}$ in the electronic ground state. However, the explicit electronic energy data are reported only
for one particular equilateral triangular configuration at the triangle size $R=1.68 \, a.u.$.
In conclusion, accurate {\em ab initio} results on the basis of which the non-existence of $\rm H_3^{++}$ can be demonstrated are scarce and not that convincing. This question is thus addressed once again in the present study.
One of the motivations of our study is related to a fact that $\rm H_3^{++}$
in equilateral triangular configuration may exist as metastable state in a magnetic field $B \gtrsim 10^8$\,G \cite{Turbiner:2002}.

\section{Methods}

\begin{figure}[htb]
\centering
\includegraphics[angle=0,
 width=0.4\textwidth]{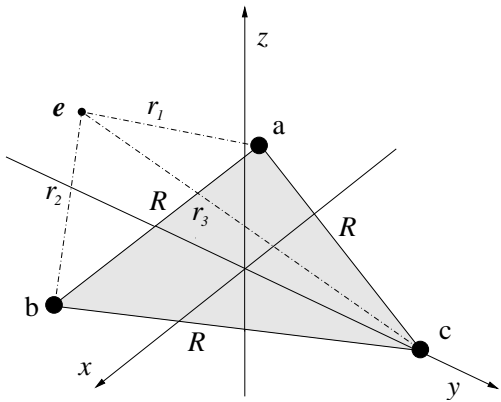}
 \caption{
   Molecular ion $\rm H_3^{++}$ in equilateral triangular
   configuration.  Protons are fixed on the vertexes (on the x-y
   plane). Point {\em{\bf e}} represents the electron position.
\label{trian}}
\end{figure}

We study a Coulomb system of one electron and three protons $(pppe)$
which form an equilateral triangle of size $R$. The protons are assumed
to be infinitely massive according to the Born-Oppenheimer approximation at zero order. The Schr\"odinger equation for the system is written as
\begin{equation}
 \left[ {\mathbf p}^2 + \frac{6}{R}-\frac{2}{r_1} -\frac{2}{r_2}-
   \frac{2}{r_3}\right]\Psi({\mathbf r}) = E\Psi({\mathbf r})  \ ,
\end{equation}
where ${\mathbf p}=-i\nabla$ is the electron momentum,
$r_1, r_2$ and $r_3$ are the distances from each proton to the electron
and $R$ is the interproton distance, see Figure~\ref{trian}.
Atomic units are used throughout ($\hbar$=$m_e$=$e$=1), although
energies are expressed in Rydbergs (Ry).

Our goal is to study the stability of the molecular ion $\rm H_3^{++}$.
If such an ion exists, it implies the existence of the ground state of the system $(pppe)$. Based on symmetry arguments it seems evident that the
optimal geometry of $(pppe)$ in the case of existence of a bound state is the equilateral triangle.

Two methods are used to explore the system: (i) variational with
physically relevant trial functions (see e.g.~\cite{Turbiner:2002}) which we will call {\it specialized} and (ii) {\it standard} variational based on using standard Gaussian trial functions as implemented
in {\em ab initio} quantum chemistry packages such as MOLPRO~\cite{MOLPRO}. Both methods lead to highly accurate quantitative results for total energy versus the size of the triangle.

In the first variational approach, a trial function is taken in a form of linear superposition of six basis functions
\begin{equation}
\label{trial}
\Psi_{trial}= \sum_{j=1}^6 A_j \psi^{(j)}\ ,
\end{equation}
where $A_j$ are linear parameters. Each function
$\psi^{(j)}$ is chosen in such a way as to describe different physical characteristics of the system. In general, $\psi^{(j)}$ has
the form of a symmetrized product of three Coulomb orbitals
\begin{equation}
\psi_C = e^{-\al_{1}r_1-\al_{2}r_2-\al_{3}r_3}\ .
\end{equation}
Let us give a brief description of each of them:

\begin{description}
\item[$\psi^{(1)}$:] All $\al$'s are chosen to be equal to $\al_1$,
\begin{equation}
  \psi^{(1)} =
  e^{-\al_{1}(r_1+r_2+r_3)}\,.
  \end{equation}
  It is a Heitler-London type function. This
  corresponds to \emph{coherent} interaction between the electron and
  all protons. Supposedly, it describes the system at small interproton
  distances and, probably, the equilibrium configuration. It might be
  verified a posteriori.

\item[$\psi^{(2)}$:] Two $\al$'s are equal to zero and the remaining one is set to be equal to $\al_2$,
\begin{equation}
  \psi^{(2)}
  =e^{-\al_{2}r_1}+e^{-\al_{2}r_2}+e^{-\al_{2}r_3} \ .
\end{equation}
    It is a Hund-Mulliken type function.
    This function possibly describes the system at large
    distances, where essentially the electron interacts with only one
    proton at a time thus realizing \emph{incoherent} interaction.

\item[$\psi^{(3)}$:] One $\al$ is equal to zero, two others
    are different from zero and equal to each other and to $\al_3$,
 \begin{equation}
  \psi^{(3)}
  =  e^{-\al_{3}(r_1+r_2)} + e^{-\al_{3}(r_1+r_3)}
           + e^{-\al_{3}(r_2+r_3)}\ .
  \end{equation}
    It is assumed that this function describes the system
    $\rm H_2^+$ plus proton when a triangle is of a sufficiently small
    size. In fact, it is the Heitler-London function of $\rm H_2^+$ symmetrized over protons.

\item[$\psi^{(4)}$:] One $\al$ is equal to zero and two others
    are different from each other being equal to $\al_{4,5}$, respectively,
\begin{eqnarray}\label{psi4}
   \psi^{(4)}
   &=&  e^{-\al_{4}r_1-\al_{5}r_2} +
   e^{-\al_{4}r_2-\al_{5}r_1} +
   e^{-\al_{4}r_1-\al_{5}r_3} \non \\
   &+&   e^{-\al_{4}r_3-\al_{5}r_1} +
   e^{-\al_{4}r_2-\al_{5}r_3} + e^{-\al_{4}r_3-\al_{5}r_2}\ .
\end{eqnarray}
    It is assumed that this function describes the system $\rm H_2^+$ plus one proton. In fact, it is the Guillemin-Zener function of $\rm H_2^+$ symmetrized over protons.
    If $\al_4=\al_5$, the function $\psi^{(4)}$ is reduced to $\psi^{(3)}$. If $\al_4=0$, the function $\psi^{(4)}$ is reduced to $\psi^{(2)}$. Hence $\psi^{(4)}$ is a non-linear interpolation between
    $\psi^{(2)}$ and $\psi^{(3)}$.  It has to describe intermediate interproton distances.

\item[$\psi^{(5)}$:] Two $\al$'s are equal but the third one is
    different,
 \begin{eqnarray}
   \psi^{(5)}
   &=&  e^{-\al_{6}r_1-\al_{6}r_2-\al_{7}r_3}
   + e^{-\al_{6}r_1-\al_{6}r_3-\al_{7}r_2}   \non\\
   &+&  e^{-\al_{6}r_2-\al_{6}r_1-\al_{7}r_3}
   + e^{-\al_{6}r_2-\al_{6}r_3-\al_{7}r_1}   \non\\
   &+&  e^{-\al_{6}r_3-\al_{6}r_1-\al_{7}r_2}
   + e^{-\al_{6}r_3-\al_{6}r_2-\al_{7}r_1}\ .
  \end{eqnarray}
    It describes a ``mixed'' state of three hydrogen atoms. If $\al_6=\al_7$, the function $\psi^{(5)}$ is reduced to $\psi^{(1)}$. If $\al_6=0$, the function $\psi^{(5)}$ is reduced to $\psi^{(2)}$. If $\al_7=0$, the function $\psi^{(5)}$ is reduced to $\psi^{(3)}$.
    Hence $\psi^{(5)}$ is a non-linear interpolation between $\psi^{(1)}$,
    $\psi^{(2)}$ and $\psi^{(3)}$. As function~(\ref{psi4}) this is a type of Guillemin-Zener function and should describe intermediate interproton distances.

\item[$\psi^{(6)}$:] All $\al$'s are different,
\begin{eqnarray}\label{psi6}
  \psi^{(6)}
  & = & e^{-\al_{8}r_1-\al_{9}r_2-\al_{10}r_3}
  + e^{-\al_{8}r_1-\al_{9}r_3-\al_{10}r_2} \non\\
  & + & e^{-\al_{8}r_2-\al_{9}r_1-\al_{10}r_3}
  + e^{-\al_{8}r_2-\al_{9}r_3-\al_{10}r_1} \non\\
  & + & e^{-\al_{8}r_3-\al_{9}r_1-\al_{10}r_2}
  + e^{-\al_{8}r_3-\al_{9}r_2-\al_{10}r_1}\ .
\end{eqnarray}
    This is a general non-linear interpolation of all functions $\psi^{(1-5)}$.
\end{description}

The total number of the parameters of the function (\ref{trial}) is equal to 15, where five of them are linear ones.
Note that $A_6$ can be fixed at $A_6 = 1$.
\\

In standard {\em ab initio} calculations, $\Psi_{trial}$ is most commonly expanded in terms of Gaussian basis
functions~\cite{BOY50:542} $\psi_{I,i}$ centered at atoms $I=1,2,3$,
\begin{equation}
\Psi_{trial} = \sum_{I=1}^3 \sum_{i=1}^n \psi_{I,i}({\bf r}_I) c_{Ii}
\end{equation}
whose coefficients $c_{Ii}$ are then determined variationally. The basis functions $\psi_{Ii}$ themselves are built up by primitive Gaussians~\cite{HUZ65:1293}
\begin{equation}
\psi_{I,i=\{n,l,m\}}({\bf r}_I) = \sum_{\kappa} N Y_{lm}(\theta,\phi) r^{2n-2-l} e^{-\alpha_{\kappa} r^2} d_{\kappa}
\end{equation}
with contraction coefficients $d_{\kappa}$ held fixed.

Our calculations were performed using the Hartree-Fock code implemented
in the MOLPRO suite of programs~\cite{MOLPRO} with the correlation consistent cc-pV6Z and modified mcc-pV7Z basis sets~\cite{DUN89:1007, MIE02:4142}.
The cc-pV6Z basis set contains 91 contracted Gaussians per atom, with $\ell$ quantum numbers up to $\ell=5$, i.e. $\rm [6s\, 5p\, 4d\, 3f\, 2g\, 1h]$,
yielding a total of 273 basis functions. The mcc-pV7Z basis includes $\ell=6$ functions, leading to 140 contracted Gaussians per atom,
$\rm [7s\, 6p\, 5d\, 4f\, 3g\, 2h\, 1i]$, or 420 basis functions in total. Calculations were carried out for a range of equilateral triangular
configurations using $C_s$ symmetry. In this point group, the lowest electronic state is $\rm ^2A^{\prime}$. The total number of contracted
Gaussians of this symmetry is 168 for the cc-pV6Z basis set and 255 for the mcc-pV7Z basis set, respectively.
The cc-pV6Z results, which are not reported here explicitly, have been generated to assess the accuracy of this type of calculations. Judging from such a comparison, we estimate the accuracy of the mcc-pV7Z data to about $10^{-5} a.u.$ over a large range of distances, deteriorating somewhat at short distances where the basis functions tend to become linearly dependent.

\section{Results}
In framework of the specialized variational method (i) some numerical computations were made. The minimization routine MINUIT~\cite{CERNLIB}
from the CERN-LIB library was used as well as D01FCF routine from the NAG-LIB~\cite{NAGLIB} for three-dimensional numerical integration.
Numerical values of the total energy $E_T$ of the system $(pppe)$ for different values of the interproton distance $R$ were obtained, see Table~\ref{tmedel}.
The results from the MOLPRO calculation with a huge standard-type basis set (mcc-pV7Z) are given for comparison. A problem of the standard
apprach is its slow convergence with respect to the angular momentum quantum number $\ell$, requiring the use of large basis sets.
The method based on the specially tailored trial function, Eq.~(\ref{trial}), leads to systematically lower variational energy values with considerably less terms. It should be noted that this method relies on a careful optimization of non-linear parameters.

\begin{table}[ht]
\centering
\begin{tabular}{|c|r|r|}
\hline
Size $R$(a.u.)&\multicolumn{2}{|c|}{Total Energy $E_T$(Ry)}\\\cline{2-3}
& Variational, specialized & Variational, standard\\
\hline
0.10& 51.2605684&51.2608280\\
0.20& 21.7652245&21.7658978\\
0.30& 12.3060338&12.3066419\\
0.40&  7.8158809& 7.8161665\\
0.50&  5.2770999& 5.2772488\\
0.60&  3.6886676& 3.6887578\\
0.70&  2.6262531& 2.6263119\\
0.80&  1.8810732& 1.8811131\\
0.90&  1.3394273& 1.3394556\\
1.00&  0.9346477& 0.9346656\\
1.10&  0.6253608& 0.6253729\\
1.20&  0.3847163& 0.3847301\\
1.30&  0.1946538& 0.1946668\\
1.40&  0.0426368& 0.0426474\\
1.50& -0.0802535&-0.0802489\\
1.65& -0.2237211&-0.2237116\\
2.00& -0.4355913&-0.4355854\\
2.50& -0.5827041&-0.5826899\\
3.00& -0.6486047&-0.6485824\\
3.50& -0.6793259&-0.6793070\\
\hline
\end{tabular}
 \caption{Variational results obtained with the specialized method (\ref{trial}) and with a standard quantum chemistry method MOLPRO for the total energy $E_T$ as a function of the internuclear distance $R$ for the system $pppe$ in the equilateral triangular geometry. For $R=3.50$, in~\cite{carbo}  $E_T=-0.678$.
\label{tmedel}}
\end{table}

Different studies have been done for $R=1.68$\, a.u. This distance corresponds to an early estimate of the equilibrium distance ($R_e$) for the molecular ion
$\rm H_3^+$ in triangular equilateral configuration~\cite{conroy2}. It provides a natural explanation why this was considered. Nowadays it is known with high
accuracy that the equilibrium distance for $\rm H_3^+$ is $R=1.65$~\cite{meyer}. We present, in Table~\ref{t168}, a comparison of the results obtained at $R=1.68$
by the standard method of quantum chemistry, LCAO, which uses optimized Gauss-type (GTO) or Slater-type (STO) atomic orbitals, the Finite Element method, the MOLPRO calculation with the massive aug-mcc-pVZ basis set~\cite{MIE02:4142} consisting of 567 basis functions,
336 of which have symmetry $A'$, and the variational approach, Eq.~(\ref{trial}).

\begin{table}[hbt]
\centering
\begin{tabular}{|l|c|l|}
\hline
  Author&Method&Energy(Ry)\\\hline
  Conroy~\cite{conroy1}&STO&-0.2431714\\\hline
  Johnson and Poshusta~\cite{johnson}& GTO&-0.24748, est. exact
  -0.248\\\hline
  Schwartz and Schaad~\cite{schwartz}& Variational& -0.24746\\\hline
  Somorjai and Yue~\cite{somorjai}& Variational& -0.23416\\\hline
  Ackermann {\it et al.}~\cite{ackermann}&F. E.&-0.2477134\\\hline
  Present work & MOLPRO, aug-mcc-pV7Z &-0.2477064\\\hline   
  Present work & Variational (\ref{trial}) &-0.2477132\\\hline
\end{tabular}
\caption{Comparison different results for $E_T$ at
  $R=1.68$\ a.u.
\label{t168}}
\end{table}

\begin{figure}[H!tb]
\centering
\psfrag{Energy}{\huge{ $E_T$(Ry)}}
\psfrag{Distance}{\huge{ $R$(a.u.)}}
\includegraphics[angle=-90, width=1.0\textwidth]{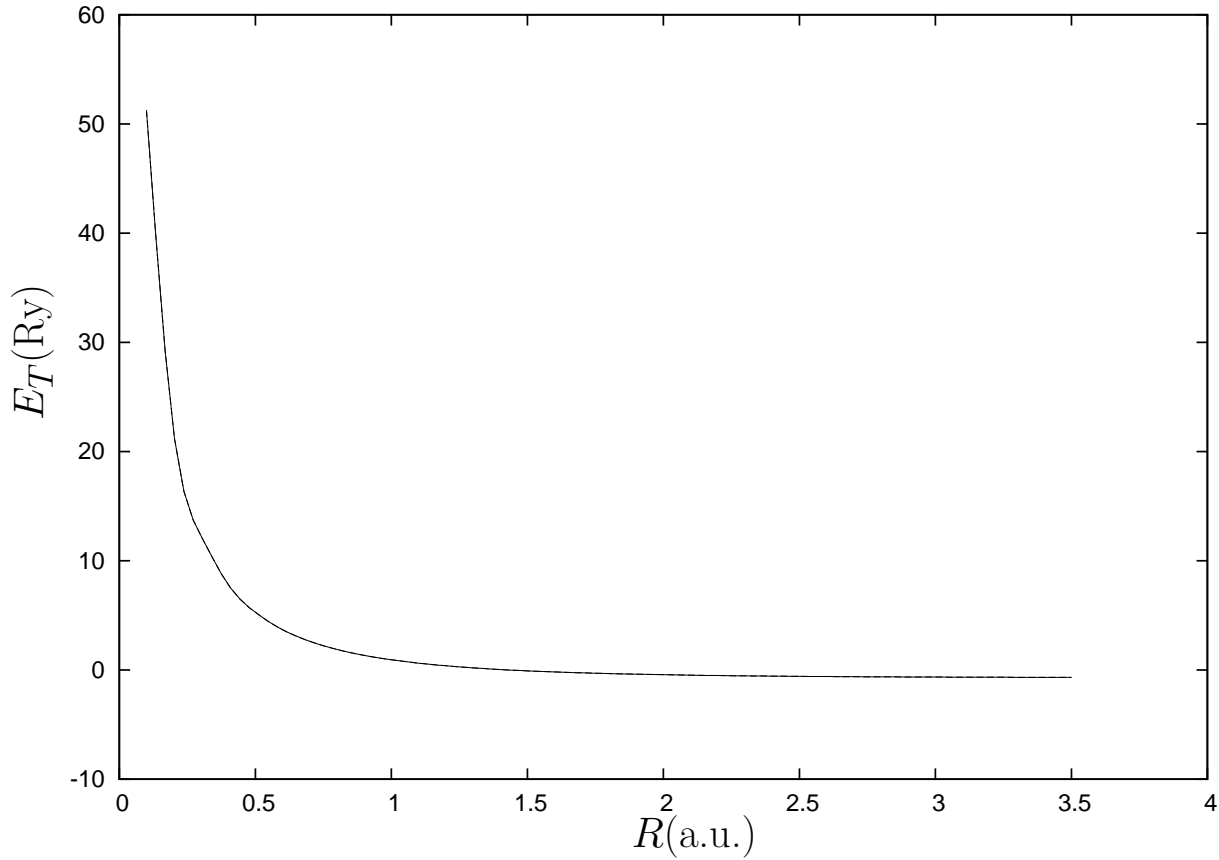}
\caption{$E_T$ as function of the internuclear distance $R$.
\label{comparasion}}
\end{figure}

Plotting the data results in a smooth monotonous curve of the total Energy $E_T$ as a function of the internuclear distance $R$, see Fig.~\ref{comparasion}.

As a conclusion we have to state that the total energy $E_T$ as a function of the internuclear distance $R$ does not indicate either to a minimum or even slight non-adiabatic irregularities for finite $R$. Thus, the molecular ion $\rm H_3^{++}$ does not exist in equilateral triangular configuration in the Born-Oppenheimer approximation.

\acknowledgements

HMC expresses his deep gratitude to J.C. L\'opez Vieyra for the valuable comments and for their interest in the present work, AVT thanks Universite Libre de Bruxelles for the hospitality extended to him where this work was completed. This work was supported in part by FENOMEC and PAPIIT grant {\bf IN121106} (Mexico).

\end{document}